\input{aipcheck}

\documentclass[
    ,final            
  ]
  {aipproc}

\layoutstyle{6x9}
\usepackage{amssymb}  
\usepackage[rightcaption]{sidecap}
\usepackage[font=footnotesize,labelfont=bf,labelsep=period]{caption}

\newcommand{\be}{\begin{equation}}
\newcommand{\ee}{\end{equation}}
\newcommand{\bea}{\begin{eqnarray}}
\newcommand{\eea}{\end{eqnarray}}

\begin{document}

\title{Chiral transition in dense, magnetized matter}


\author{Florian Preis}{address={Institut f\"ur Theoretische Physik, Technische Universit\"at Wien, 1040 Vienna, Austria}}

\author{Anton Rebhan}{
  address={Institut f\"ur Theoretische Physik, Technische Universit\"at Wien, 1040 Vienna, Austria}
}

\author{Andreas Schmitt\footnote{speaker}}{
  address={Institut f\"ur Theoretische Physik, Technische Universit\"at Wien, 1040 Vienna, Austria} 
}

\begin{abstract}

In the presence of a chemical potential, the effect of a magnetic field on chiral symmetry breaking goes beyond the well-known
magnetic catalysis. Due to a subtle interplay with the chemical potential, the magnetic field may work not only in favor but also against the chirally broken phase. 
At sufficiently large coupling, the magnetic field favors the broken phase only for field strengths beyond any conceivable value in nature. 
Therefore, in the interior of magnetars, a possible 
transition from chirally broken hadronic matter to chirally symmetric quark matter might occur at smaller densities than  previously thought.

\end{abstract}

\maketitle


\section{Introduction}

Spontaneous breaking of (approximate) chiral symmetry is a fundamental, non-perturbative aspect of quantum chromodynamics (QCD). Within lattice gauge theory, 
the chiral condensate can be
computed, and it has been shown that QCD exhibits a smooth transition (``crossover'') from the chirally broken to the chirally restored phase \cite{Aoki:2006we}. 
This crossover occurs at $T\simeq 150\, {\rm MeV}$. Little is known from first principles about the transition at nonzero chemical potentials $\mu$ 
and/or magnetic fields $B$, the exception being asymptotically large values of $\mu$, where chiral symmetry is broken by the formation of a diquark condensate in the 
color-flavor 
locked (CFL) phase \cite{Alford:1998mk,Alford:2007xm}. At $\mu=0$, lattice studies of the chiral phase transition in a magnetic field have just begun recently, and 
first results indicate the decrease of the crossover temperature with increasing $B$ \cite{Bali:2011qj,D'Elia:2012tr}. 
From phenomenological models, for instance the Nambu-Jona-Lasinio (NJL) model, one gets indications that can be summarized as follows,
\begin{itemize}

\item $\mu=0$, $B\neq 0$. ``Magnetic catalysis'' \cite{Klimenko:1990rh,Gusynin:1994re}: 
the magnetic field enhances the chiral condensate, the critical temperature is expected to increase 
with $B$. (Note that this expectation is not borne out by the recent lattice results. For possible explanations of this discrepancy see 
Refs.\ \cite{Miransky:2002rp,Fukushima:2012kc}.) 

\item $B=0$, $\mu\neq 0$. Chiral symmetry is expected to be restored at sufficiently large $\mu$, most likely in a first-order phase transition. (If the 
hadronic phase is directly superseded by CFL, a crossover is conceivable at small temperatures \cite{Hatsuda:2006ps,Schmitt:2010pf}.)   

\end{itemize}
How about the situation where both $\mu$ and $B$ are nonvanishing? Naively extrapolating the above two indications suggests a critical chemical potential
that increases with the magnetic field. Indeed, this is found for weak coupling in the NJL model. In the following, we discuss that at strong coupling this expectation
is incorrect because of a nontrivial interplay between the magnetic field and the chemical potential. The resulting effect is called ``inverse magnetic
catalysis'' \cite{Preis:2010cq}.

\section{Magnetic Catalysis}

Magnetic catalysis can be explained with the help of an NJL model for one flavor of massless fermions,
\be
 \mathcal{L}=\overline{\psi}(i\gamma^\mu \partial_\mu+\mu\gamma_0)\psi+G\left[\left(\overline{\psi}\psi\right)^2 
+\left(\overline{\psi}\gamma_5 \psi\right)^2\right] \, ,
\ee
where $G$ is a coupling constant with mass dimensions -2. Within the mean-field approximation, one can show that a dynamical fermion mass
$M\equiv -2G\langle\overline{\psi}\psi\rangle$ is generated only for sufficiently large coupling strengths $g>1$, where $g\equiv G\Lambda^2/(2\pi)$ with an ultraviolet
momentum cut-off $\Lambda$. (The precise numerical value of the critical dimensionless coupling, here $g=1$, depends on the regularization scheme; the results we are 
referring to here \cite{Preis:2012fh}
are done within the proper time regularization where $1/\Lambda^2$ is the lower bound of the proper time integral.)  

If the same analysis is done in the presence of a background magnetic field $B$, the conclusion is changed qualitatively. Now, a dynamical mass is generated
for {\it all} (i.e., arbitrarily small) coupling strengths. The weak-coupling result ($g\ll 1$) is 
\be \label{M}
M \simeq \sqrt{\frac{B}{\pi}}\,e^{-\frac{\pi^2}{BG}} \, .
\ee
(For simplicity, we have set the electric charge of the fermions to 1.) The physics behind this effect can be nicely understood in analogy to BCS theory: for an 
arbitrarily weak attractive interaction and sufficiently small temperatures, any Fermi surface is unstable with respect to the formation of a Cooper
pair condensate. More technically speaking, an infrared divergence is induced by the Fermi surface because it renders the system effectively 1+1 dimensional.
Here, the magnetic field plays the role of the Fermi surface in some sense because, due to the physics of the lowest Landau level (LLL), the system also becomes
effectively 1+1 dimensional. As a consequence, an instability analogous to the Cooper instability arises, and a fermion--anti-fermion condensate is formed. The 
exponential form of the dynamical mass (\ref{M}) is thus analogous to the weak-coupling gap in a BCS superconductor. 

Magnetic catalysis manifests itself also in the critical temperature $T_c$ for chiral symmetry breaking: within the NJL model, one finds that $T_c$ increases 
monotonically with increasing magnetic field. The result for large $B$ suggests that $T_c$ can become arbitrarily large (although the NJL model does not allow for reliable
predictions for magnetic fields beyond the cutoff scale). 

Before we come to nonzero chemical potentials, let us, in addition to the NJL model, also introduce the Sakai-Sugimoto model \cite{Sakai:2004cn}, which is 
a specific example of the gauge/gravity duality. The Sakai-Sugimoto model is based on type-IIA string theory, and is currently the holographic model whose dual comes
closest to QCD. This is achieved by breaking supersymmetry through the
compactification of an extra dimension on a circle, which
gives mass to unwanted adjoint scalars and fermions of
the order of the Kaluza-Klein mass $M_{\rm KK}$ (inversely
proportional to the radius of the compactification circle).
Moreover, it realizes spontaneous symmetry breaking of the full chiral group $SU(N_f)_R\times SU(N_f)_L$ in a geometrical way
with the help of $N_f$ D8 and $\overline{\rm D8}$ branes, embedded in the background of $N_c$ D4 branes. 
In its original version, chiral symmetry breaking is rigidly coupled to the deconfinement phase transition, i.e., chiral symmetry is restored if and only if 
the deconfined phase is preferred. The chirally restored, deconfined phase occurs above a certain critical temperature which is independent of the values of $\mu$ and $B$.
The reason 
for this trivial phase structure is the large-$N_c$ limit, to which the usually employed probe brane approximation is necessarily restricted. However, besides  
$M_{\rm KK}$ there is a second parameter $L$ in the model (the asymptotic separation of the flavor D8 and $\overline{\rm D8}$ branes in the compactified direction) 
which can be varied to obtain a less rigid behavior. 
The reason is that for small $L$ the gluon dynamics become less important. Effectively, confinement is switched off, and thus for $L\ll M_{\rm KK}^{-1}$ 
the model can be expected to be dual to an NJL-like field theory. Although its physical content is less transparent than the one of the NJL  model, 
it is a very useful tool since it yields a top-down, microscopic description of strongly coupled physics. 

Within the Sakai-Sugimoto model, the usual magnetic catalysis is confirmed: the critical temperature as well as the constituent quark mass increase monotonically 
with $B$ \cite{Johnson:2008vna}. In contrast to the NJL model, $T_c$ saturates at a finite value for asymptotically large $B$, i.e., in some sense magnetic catalysis 
becomes weaker for large $B$.   

\section{Inverse Magnetic Catalysis}

To discuss the chiral phase transition in the presence of nonzero $B$ {\it and} $\mu$, we consider the free energy difference $\Delta\Omega$ between the 
chirally restored and chirally broken phases, such that, by convention, the ground state is chirally restored for $\Delta\Omega>0$ and chirally broken for 
$\Delta\Omega<0$. First, we are interested in the weak-coupling limit. This is easily done in the NJL model. In the Sakai-Sugimoto model, we are restricted to the 
strongly coupled limit. Interestingly, however, we recover the weak-coupling NJL result when we take the limit of large magnetic fields in the holographic 
calculation. Hence we can write for both models,
\be \label{MC}
\Delta\Omega =  \frac{B}{4\pi^2}[\mu^2-\alpha M(B)^2] \, ,
\ee
where $\alpha = \frac{1}{2}$ (weak-coupling NJL) and $\alpha= \sqrt{\pi}\Gamma\left(\frac{3}{5}\right)/\left[3\Gamma\left(\frac{1}{10}\right)\right]\simeq 0.09$ 
(large-$B$ 
Sakai-Sugimoto\footnote{For a better comparison, here we have omitted the effect of the topologically induced baryon density (``chiral spiral''). This effect is included
in the Sakai-Sugimoto phase diagrams of Fig.\ \ref{figSS} and does, within this particular model, not change the chiral phase transition qualitatively.}). 
This expression for 
$\Delta\Omega$ is in complete accordance with magnetic catalysis: if we start from the chirally broken phase, i.e., $\Delta\Omega<0$, we can never restore chiral
symmetry by increasing $B$ at fixed $\mu$ because the constituent quark mass $M(B)$ increases with $B$. 

\begin{SCfigure}[\sidecaptionrelwidth]
\includegraphics[height=5cm]{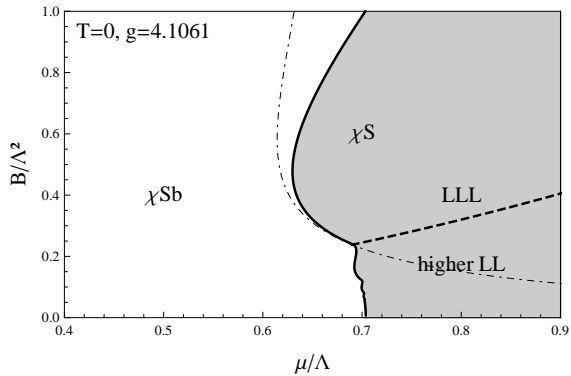}
\vspace{-1cm}
\caption{Zero-temperature phase diagram in the $\mu$-$B$ plane within the NJL model. Due to the large coupling chosen here, the chiral phase transition (solid line)
is first order. Up to $B/\Lambda^2\lesssim 0.5$ the critical chemical potential decreases with $B$ (inverse magnetic catalysis). 
The dashed-dotted line is the approximation from Eq.\ (\ref{IMC}) and shows that inverse magnetic catalysis is mainly a LLL phenomenon.}
\label{figNJL}
\vspace{.3cm}
\end{SCfigure}

\begin{figure}
\hbox{ \includegraphics[width=.5\textwidth]{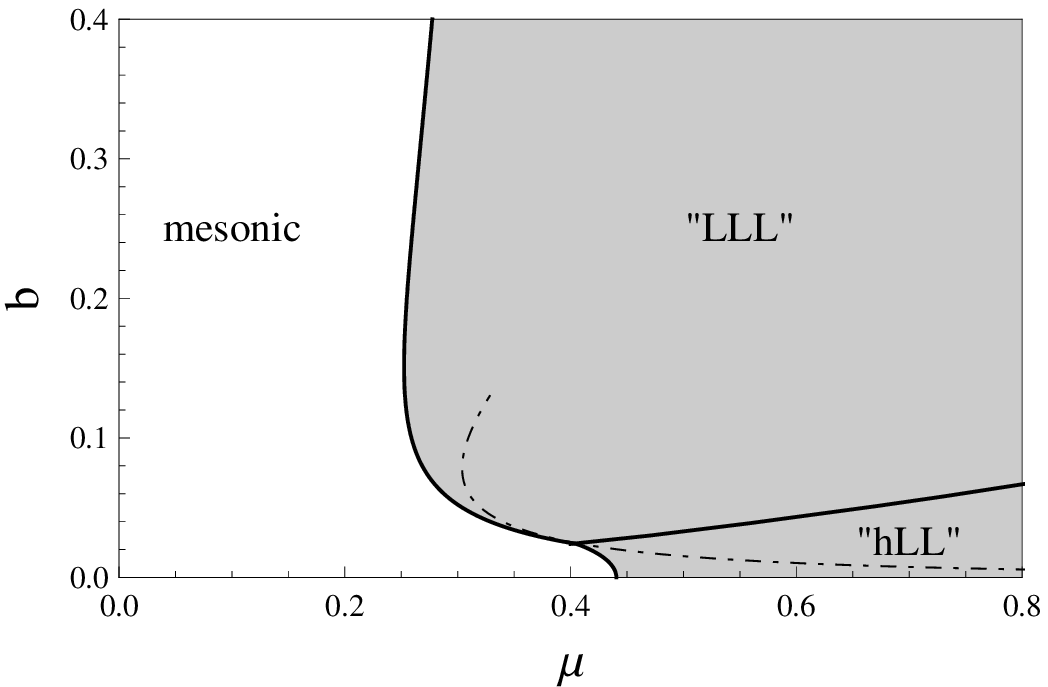}\includegraphics[width=.5\textwidth]{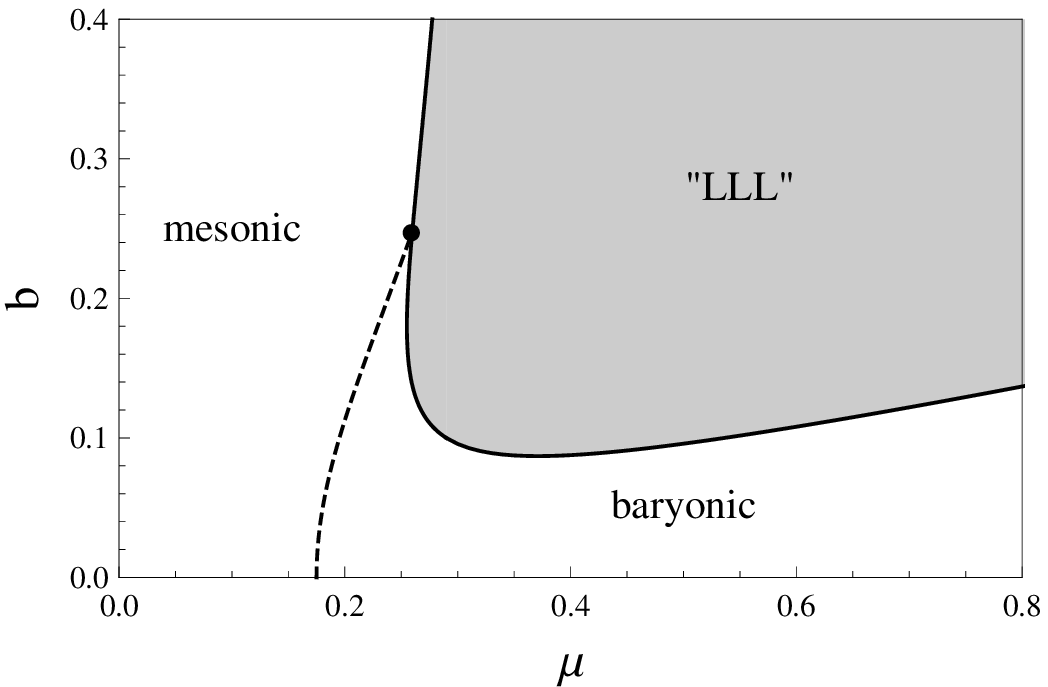}}
  \caption{Left panel: zero-temperature chiral phase transition in the Sakai-Sugimoto model, ignoring baryonic matter. The phase diagram is very similar
to the NJL result in Fig.\ \ref{figNJL}. Again, the dashed-dotted line is obtained from Eq.\ (\ref{IMC}). Right panel: taking into account baryonic matter changes 
the phase diagram dramatically. The baryon onset line (dashed) intersects the chiral phase transition line. For small magnetic fields, chiral symmetry remains 
spontaneously broken for arbitrarily large chemical potentials. The dimensionless
scales can be translated into physical ones assuming that the phase transition without baryons at $b=0$ occurs at $\mu\sim 400\,{\rm MeV}$. Then, $b=0.1$ corresponds to 
a magnetic field $B\simeq 1.0\times 10^{19}\,{\rm G}$ \cite{Preis:2010cq}.}
\label{figSS}
\end{figure}

For supercritical couplings $g>1$ in the NJL model the situation is different. In this regime, the small-$B$ limit of the NJL model is very similar to the 
small-$B$ limit of the Sakai-Sugimoto model (where the coupling is always large). For simplicity, we only consider the LLL. This is also possible
in the Sakai-Sugimoto model, where a structure reminiscent of the LLL has been observed \cite{Preis:2010cq,Lifschytz:2009sz}. 
The LLL approximation is of course inconsistent with the limit of very small magnetic fields. Nevertheless, we find that there is an intermediate regime 
where this approximation reproduces the full numerical result, and it captures the essential interplay between $\mu$ and $B$. We find
\be \label{IMC}
\Delta\Omega =  -(a_0 + a_1 B^2) + \frac{B\mu^2}{4\pi^2} \, ,
\ee
with the positive constants $a_0\propto M_0^2\Lambda^2$ (NJL), $a_0\propto M_0^{7/2} M_{\rm KK}^{1/2}$ (Sakai-Sugimoto), where $M_0$ is the constituent quark mass 
in the absence of a magnetic field. Also $a_1$ is positive, for the explicit expressions see
Ref.\ \cite{Preis:2012fh}. Now, the chiral phase transition does not behave as naively expected from magnetic catalysis. Even though the chiral
condensate still increases with $B$, $\Delta\Omega$ can switch its sign upon increasing $B$. The reason is that the condensation energy only increases quadratically 
with $B$, while the positive term in the free energy difference is linear in $B$. As a consequence, a magnetic field can now restore chiral symmetry at a given chemical 
potential. This effect is called inverse magnetic catalysis and is confirmed by the full numerical results which are shown in Figs.\ \ref{figNJL} (NJL) and 
\ref{figSS} (Sakai-Sugimoto), including all Landau levels. In the latter figure, we also show the chiral phase transition in the presence of homogeneous baryonic 
matter \cite{Preis:2011sp}.

Inverse magnetic catalysis is thus a manifestation of a somewhat ambivalent role of the magnetic field. It enhances the condensation energy, but
also leads to a larger free energy cost which, in the presence of a chemical potential, has to be paid for fermion--anti-fermion pairing. It is instructive
to compare this situation with Cooper pairing in the presence of a mismatch of the Fermi momenta of the constituent fermions. In this
case, one can imagine the condensation process by first filling the two fermion species up to a common 
Fermi momentum, and then forming the pairs \cite{Rajagopal:2005dg}. The first step costs free energy depending on the mismatch, the second yields a gain, 
depending on the condensation energy (usually given by the coupling strength). 
Our situation is analogous, as Eqs.\ (\ref{MC}) and (\ref{IMC}) illustrate: in the first step we need to overcome the mismatch
of fermions and antifermions, in the second we form the chiral condensate. Now, crucially, the change in free energy of {\it both} steps depend on $B$. Unless this 
dependence is the same, as in Eq.\ (\ref{MC}),  inverse magnetic catalysis becomes possible.

\bibliographystyle{aipproc}   

\end{document}